\RequirePackage{etoolbox}\csdef{input@path}{{style/}{graphics/}}
\documentclass[aoas,MSNbibl,nameyear,dvips]{arximspdf}
\usepackage{graphicx}
\doi{10.1214/13-AOAS687} 
\volume{8}
\issue{1}
\pubyear{2014}
\firstpage{309}
\lastpage{330}

\makeatletter
\def\mid{|}
\newcommand{\bnu}{{\bolds{\nu}}}
\newcommand{\bpi}{{\bolds{\pi}}}
\makeatother

\begin{document}
\begin{frontmatter}

\title{Quantifying alternative splicing from paired-end RNA-sequencing data}
\runtitle{Alternative splicing from RNA-seq data}

\begin{aug}
\author[a]{\fnms{David} \snm{Rossell}\corref{}\ead[label=e1]{D.Rossell@warwick.ac.uk}\thanksref{t1,m1}}
\author[b]{\fnms{Camille} \snm{Stephan-Otto Attolini}\ead[label=e2]{camille.stephan@irbbarcelona.org}\thanksref{tt1,m2}}\break
\author[c]{\fnms{Manuel} \snm{Kroiss}\ead[label=e3]{kroissm@in.tum.de}\thanksref{m3,m4,ttt1}}
\and
\author[d]{\fnms{Almond} \snm{St\"ocker}\ead[label=e4]{al.st@web.de}\thanksref{m3,ttt1}}
\thankstext{t1}{Supported in part by Grants R01 CA158113-01 from the
NIH (USA)
and MTM2012-38337 from the Ministerio de Econom\'ia y Competitividad (Spain).}
\thankstext{tt1}{Supported in part by AGAUR Beatriu de Pin\'os
fellowship BP-B 00068 (Spain).}
\thankstext{ttt1}{Supported in part by a Bayerisches Arbeitsministerium
foreign trainee fellowship (Germany).}
\runauthor{Rossell, Stephan-Otto Attolini, Kroiss and St\"ocker}
\affiliation{University of Warwick\thanksmark{m1},
Institute for Research in Biomedicine of Barcelona\thanksmark{m2}, LMU
Munich\thanksmark{m3} and TU Munich\thanksmark{m4}}

\address[a]{D. Rossell\\
Department of Statistics\\
University of Warwick\\
Gibbel Hill Rd.\\
Coventry CV4 7AL\\
United Kingdom\\
\printead{e1}}

\address[b]{C. Stephan-Otto Attolini\\
Institute for Research in Biomedicine\\
\quad of Barcelona\\
Baldiri Reixac 10\\
Barcelona 08028\\
Spain\\
\printead{e2}}

\address[c]{M. Kroiss\\
LMU Munich\\
Geschwister-Scholl-Platz 1\\
M\"unchen 089 2180-0\\
Germany\\
\printead{e3}}

\address[d]{A. St\"ocker\\
TU Munich\\
Geschwister-Scholl-Platz 1\hspace*{38pt}\\
M\"unchen 089 2180-0\\
Germany\\
\printead{e4}}

\end{aug}

\received{\smonth{9} \syear{2012}}
\revised{\smonth{5} \syear{2013}}

%
\begin{abstract}
RNA-sequencing has revolutionized biomedical research and, in
particular, our ability to study gene alternative splicing. The
problem has important implications for human health, as alternative
splicing may be involved in malfunctions at the
cellular level and multiple diseases. However, the high-dimensional
nature of the data and
the existence of experimental biases pose serious data analysis challenges.
We find that the standard data summaries used to study alternative splicing
are severely limited,
as they ignore a substantial amount of valuable information.
Current data analysis methods are based on such summaries
and are hence suboptimal.
Further, they have limited flexibility in accounting for technical biases.
We propose novel data summaries and a Bayesian modeling framework that
overcome these
limitations
and determine biases in a nonparametric, highly flexible manner.
These summaries adapt naturally to the rapid improvements in sequencing
technology.
We provide efficient point estimates
and uncertainty assessments. The approach
allows to study alternative splicing patterns for individual
samples and can also be the basis for downstream analyses.
We found a severalfold improvement in estimation mean square error
compared popular approaches in simulations,
and substantially higher consistency between replicates in experimental data.
Our findings indicate the need for adjusting the routine summarization
and analysis of alternative splicing RNA-seq studies.
We provide a software implementation in the R package \texttt{casper}.%
 \def\thefootnote{4}\footnote{\href{http://www.bioconductor.org/packages/release/bioc/html/casper.html}{http://www.bioconductor.org/packages/release/bioc/html/casper.html}.}
\end{abstract}

%

\end{frontmatter}

\section{Introduction}
\label{sec:intro}

RNA-sequencing (RNA-seq) produces an overwhelming amount of genomic
data in a single experiment, providing an unprecedented resolution to
address biological problems. We focus on gene expression experiments
where the goal is to study alternative splicing (AS), which we briefly
introduce. AS is an important biological process by which cells are
able to express several variants, also known as isoforms, of a single
gene. Each splicing variant gives rise to a different protein with a
unique structure that can perform different functions and respond to
internal and environmental needs. AS is believed to contribute to the
complexity of higher organisms, and is in fact particularly common in
humans [\citet{Blencowe:2006}]. Additionally, it is known to be
involved in multiple diseases such as cancer and malfunctions at the
cellular level. Despite its importance, due to limitations in earlier
technologies, most gene expression studies have ignored AS and focused
on overall gene expression.

\begin{figure}

\includegraphics{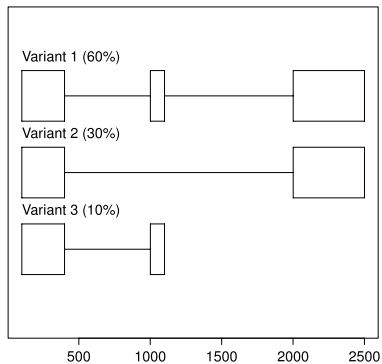}

\caption{Three splice variants for a hypothetical gene and their
relative abundances. Exon 1 is located at positions 101--400. Exon 2 at
1001--1100. Exon 3 at 2001--2500.}
\label{fig:toygene}
\end{figure}

Consider the hypothetical example of a gene with three splice variants
shown in Figure~\ref{fig:toygene}.
The gene is encoded in the DNA in three exons, shown as boxes in
Figure~\ref{fig:toygene}. When the gene is transcribed as messenger
RNA (mRNA), it can give rise to three isoforms. Variant 1 is formed
by all three exons, whereas variant 2 skips the second exon and
variant 3 the third exon. Usually, multiple variants are expressed
simultaneously at any given time. In our example, variant 1 makes up
for 60\% of the overall expression of the gene, variant 2 for 30\% and
variant 3 for 10\%. In practice, these proportions are unknown and
our goal is to estimate them as accurately as possible.

We focus on paired-end RNA-seq experiments, as they
are the current standard and provide higher
resolution for measuring isoform expression than competing
technologies, {for example}, microarrays [\citet{pepke:2009}]. RNA-seq
sequences tens or even hundreds of millions of mRNA fragments, which can
then be aligned to a reference genome using a variety of software,
{for example}, TopHat [\citet{trapnell:2009}], SOAP [\citet
{liSoap:2009}] or
BWA [\citet{liBwa:2009}]. Throughout, we assume that the software can
handle gapped alignments (we used TopHat in all our examples). Early
RNA-seq studies used single-end sequencing, where only the left or
right end of a fragment is sequenced.
In contrast, paired-end RNA-seq sequences both fragment ends.
Table~\ref{tab:toydata} shows three hypothetical
sequenced fragments corresponding to the gene in Figure~\ref{fig:toygene}. 75 base pairs (bp) were sequenced from each end.
For instance, both ends of fragment 1 align to exon 1. As the three
variants contain exon 1, in principle, this fragment could have been
generated by any variant. For fragment 2 the left read aligned to
exons 1 and 2 ({i.e.}, it spanned the junction between both
exons), and the right read to exon 3. Hence, fragment 2 can only have
been generated from variant 1. Finally, fragment 3 visits exons 1 and
2 and, hence, it could have been generated either by variants 1 or 3. The
example is simply meant to provide some intuition. In practice, most
genes are substantially longer and have more complicated splicing
patterns. Precise probability calculations
are required to ensure that the conclusions are sound.

\begin{table}
\caption{Three paired-end RNA-seq fragments.
Aligned chromosome and base pairs are indicated for both ends,
allowing for gapped alignments.
The exon path indicates the sequence of exons visited by each end.
A typical experiment contains tens of millions of fragments}
\label{tab:toydata}
\begin{tabular*}{\textwidth}{@{\extracolsep{\fill}}lcccc@{}}
\hline
& \textbf{Chromosome} & \textbf{Left read} & \textbf{Right read} & \textbf{Exon path} \\
\hline
Fragment 1 & chr1 & 110--185 & 200--274 & \{1\}, \{1\} \\
Fragment 2 & chr1 & 361--400; 1001--1035 & 2011--2085 & \{1, 2\}, \{3\} \\
Fragment 3 & chr1 & 301--375 & 1021--1095 & \{1\}, \{2\} \\
$\cdots$ & & & & \\
\hline
\end{tabular*}
\end{table}

Ideally, one would want to sequence the whole variant,
so that each fragment can be uniquely assigned to a variant.
Unfortunately, current technologies sequence hundreds of base pairs,
which are orders of magnitude shorter than typical variant lengths.
Current statistical approaches are based on the observation that,
while most sequenced fragments cannot be uniquely assigned to a variant,
it is possible to make probability statements.
For instance, fragment 3 in Table~\ref{tab:toydata}
may have originated either from variant 1 or 3,
but the probability that each variant generates such a fragment is different.
As we shall see below, this observation prompts a direct use of Bayes theorem.

In principle, one could formulate a probability model that uses the
full data, {that is}, the exact base pairs covered by each fragment
such as provided in
Table~\ref{tab:toydata},
{for example}, \citet{glaus:2012}.
However, our findings indicate that such strategies can
be computationally prohibitive and deliver no obvious improvement
(Section~\ref{sec:results}).
Further, data storage and transfer requirements impose a need for
reducing the size of the data.
Several authors proposed
summarizing the data by counting the number of fragments either
covering each exon or each exon junction [{e.g.},
\citet{xing:2006}, \citet{mortazavi:2008}, \citet{jiang:2009}].
In fact, large-scale genomic databases report precisely these summaries,
{for example}, The Cancer Genome Atlas
project.\setcounter{footnote}{4}\footnote{\href{http://cancergenome.nih.gov}{http://cancergenome.nih.gov}.}
One can then pose a probability model that uses count data from a few
categories as raw data, which greatly simplifies computation. While
useful, this approach is seriously limited to considering pairwise junctions,
which discards relevant information.
For instance, suppose that a fragment visits exons 1, 2 and 3. Simply
adding 1 to the count of fragments spanning exons 1--2 and 2--3 ignores
the joint information that a single fragment visited 3 exons
and decreases the confidence when inferring the
variant that generated the fragment.
Our results suggest that ignoring this information can result
in a serious loss of precision.
It is not uncommon that a
fragment spans more than 2 exons. \citet{holt:2008} found a
substantial proportion of fragments bridging several exons in
paired-end RNA-seq experiments. In the 2009 RGASP experimental data set
(Section~\ref{sec:results}) 38.0\% and 40.9\% of fragments spanned
$\geq$3 exons in replicate 1 and 2, respectively
(we subdivided exons so that they are fully shared/not shared by all
variants in a gene).
In the 2012 ENCODE data set we found 64.7\% and 65.2\% in each replicate.
The 2012 data had substantially longer reads and fragments,
which illustrates the rapid advancements in technology.
As sequencing evolves, these percentages are expected to increase further.

We propose novel data summaries that
preserve most information relevant to alternative splicing,
while maintaining the computational burden at a manageable level.
We record the sequence of exons visited by each fragment end, which we
refer to as \textit{exon path},
and then count the number of fragments following each exon path.
The left end of Fragment 2 in Table~\ref{tab:toydata} visits exons 1
and 2
and the right end exon 3, which we denote as $\{1,2\},\{3\}$.
Notice that a fragment following the path $\{1\},\{2,3\}$ visits the
same exons,
so one could be tempted to simply record $\{1,2,3\}$ in both cases.
However, the probability of observing $\{1,2\},\{3\}$ for a given variant
differs from $\{1\},\{2,3\}$
and, hence, combining the two paths would result in a potential loss of
information.
Table~\ref{tab:pathcounts} contains hypothetical exon path counts
for our example gene.
We use these counts as the basic input for our probability model.

\begin{table}[b]
\caption{Exon path counts for hypothetical gene}
\label{tab:pathcounts}
\begin{tabular*}{150pt}{@{\extracolsep{\fill}}lc@{}}
\hline
\textbf{Exon path} & \textbf{Count} \\
\hline
\{1\}, \{1\} & 2824 \\
\{2\}, \{2\} & \phantom{0}105 \\
\{3\}, \{3\} & 5042 \\
\{1\}, \{2\} & \phantom{00}27 \\
\{1\}, \{1, 2\} & \phantom{0}423 \\
\{1\}, \{3\} & \phantom{0}127 \\
\{2, 3\}, \{3\} & \phantom{0}394 \\
\{1, 2\}, \{3\} & \phantom{000}2 \\
\{1\}, \{2, 3\} & \phantom{00}13 \\
\hline
\end{tabular*}
\end{table}

Paired-end RNA-seq is critical for AS studies. Intuitively, compared
to single-end sequencing, it increases the probability of observing
fragments that connect exon junctions. \citet{lacroix:2008} showed
that, although neither protocol guarantees the existence of a unique
solution, in practice, paired-end (but not single-end) can provide
asymptotically correct estimates for 99.7\% of the human genes. In
contrast, for single-end data the figure is 1.14\%. Unfortunately,
much of the current methodology has been designed with single-end data
in mind. \citet{xing:2006} formulate the problem as that of
traversing a directed acyclic graph and formulate a latent variable
based approach to estimate splice variant expression.
\citet{jiang:2009} propose a similar approach within the Bayesian
framework. Both
approaches were designed for single-end RNA-seq data.
\citet{ameur:2010} proposed strategies to
detect splicing junctions, and \citet{katz:2010} and
\citet{wuSpliceTrap:2011} introduced models to estimate the percentage
of isoforms skipping individual exons. However, these approaches do
not estimate expression at the variant level.

Several authors propose strategies that use paired-ends.
\citet{mortazavi:2008}, \citet{montgomery:2010}, \citet{trapnell:2010}
and \citet{salzman:2011} model the
number of fragments spanning exon junctions.
These approaches focus on
pairwise exon connections, ignoring valuable higher-order information,
and have limitations in incorporating important technical biases.
First, the
sample preparation protocols usually induce an enrichment toward the
3' end of the transcript, {that is}, fragments are not uniformly
distributed along the gene.
\citet{roberts:2011b}, \citet{wu:2011} or \citet{glaus:2012} relax the
uniformity assumption.
Further, the fragment
length distribution plays an important role in the probability calculations
and needs to be estimated accurately.
While the approaches above acknowledge this issue,
they either use sequencing facility reports ({i.e.}, they do not
estimate the distribution from the
data) or they impose strong parametric assumptions.
Our examples illustrate that facility
reports can be inaccurate and that parametric forms do not capture the observed
asymmetries, heavy tails or multi-modalities.
Further, all previous approaches assume that fragment start and length
distributions
are constant across all genes.
We provide empirical evidence that this assumption can be flawed
and suggest a strategy to relax the assumption.

A concern with current genome annotations is that they may miss some
splicing variants.
Our approach can be combined with methods that predict new variants
such as Cufflinks RABT module [\citet{trapnell:2010,roberts:2011}],
Scripture [\citet{guttman:2010}] or
SpliceGrapher [\citet{rogers:2012}].
This option is implemented in our R package and illustrated in Section~\ref{ssec:encode}.

In summary, we propose a flexible framework to estimate alternative
splicing from RNA-seq studies,
by using novel data summaries and accounting for experimental biases.
In Section~\ref{sec:model}
we formulate a probability model that goes beyond pairwise connections
by considering exon paths.
We model the read start and fragment size distributions nonparametrically
and allow for separate estimation within subsets of genes with similar
characteristics.
Section~\ref{sec:fitting} discusses model fitting and provides algorithms
to obtain point estimates, asymptotic credibility intervals and
posterior samples.
We show some results in Section~\ref{sec:results} and provide
concluding remarks in Section~\ref{sec:discussion}.

\section{Probability model}
\label{sec:model}

We formulate the model at the gene level
and perform inference separately for each gene.
In some cases, exons from different genes overlap with each other.
When this occurs 
we group the overlapping genes and consider all their isoforms
simultaneously.
It is also possible that two variants share only a part of an exon.
We subdivide such exons into the shared part and the part that is specific
to each variant.
For simplicity, from here on we refer to gene groups simply as genes
and to subdivided exons simply as exons.

Consider a gene with $E$ exons starting at base pairs
$s_1,\ldots,s_E$ and ending at $e_1,\ldots,e_E$.
Denote the set of splicing variants under consideration by $\bnu$
(assumed to be known) and its cardinality by $|\nu|$.
Each variant 
is characterized by an increasing sequence of natural numbers
$i_1,i_2,\ldots$
that indicates the exons contained therein.

\subsection{Likelihood and prior}
\label{ssec:likelihood}

As discussed above, we formulate a model for exon paths.
Let $k$ be the number of exons visited by the left read,
and $k'$ be that for the right read
({i.e.}, $k=k'=1$ when both reads overlap a single exon).
We denote an exon path by $\bolds{\iota}=(\bolds{\iota}_l,\bolds{\iota}_r)$,
where $\bolds{\iota}_l=(i_j,\ldots,i_{j+k})$ 
are the exons visited by the left-end
and $\bolds{\iota}_r=(i_{j'},\ldots,i_{j'+k'})$ those by the right-end.
Let $\mathcal{P^*}$ be the set of all possible exon paths and
$\mathcal{P}$ be the subset of observed paths,
{that is}, the paths followed by at least 1 sequenced fragment.

The observed data is a realization of the random variable
$\mathbf{Y}=(Y_1,\ldots,Y_N)$, where $N$ is the number of paired-end reads and
$Y_i \in\{1,\ldots,\mathcal{P}^*\}$ indicates the exon path followed
by read pair $i$.
Formally, $Y_i$ arises from a mixture of $|\nu|$ discrete probability
distributions,
each component corresponding to a different splicing variant.
The mixture weights $\bpi=(\pi_1,\ldots,\pi_{|\nu|})$ give the
proportion of reads generated by each variant,
{that is}, its relative expression.
That is,
\[
P(Y_i=y_i| \bpi, \bnu)= \sum
_{d=1}^{|\nu|} p_{y_id} \pi_d,
\]
where
$p_{kd}= P(Y_i=k|\delta_i=d)$ is the probability of path $k$
under variant $d$
and $\delta_i$ is a latent variable indicating the variant that
originated $Y_i$.
Let $S_i$ and $L_i$ denote the relative start and
length (resp.) of fragment $i$. The exon path $Y_i$ is
completely determined given $S_i$, $L_i$ and the variant $\delta_i$.
Hence,
%
%
\begin{equation}
p_{kd} = \int\!\!\int
 I(Y_i=k \mid
S_i=s_i, L_i=l_i,
\delta_i=d) \,dP_{L}(l_i \mid
\delta_i) \,dP_S(s_i \mid
\delta_i, L_i), \label{eq:pathprobs}
\end{equation}
where $P_L$ is the fragment distribution and
$P_S$ is the read start distribution conditional on $L$.
As discussed in Section~\ref{ssec:distrib},
by assuming that $P_S$ and $P_L$
are shared across sets of genes with similar characteristics,
it is possible to estimate them with high precision.
Hence, for practical purposes we can treat $p_{kd}$ as known
and pre-compute them before model fitting.
Full derivations for $p_{kd}$ are provided in Appendix \ref{sec:path_prob}.

Assuming that each fragment is observed independently,
the likelihood function can be written as
%
%
\begin{equation}
P(\mathbf{Y}| \bpi,\bnu)= \prod_{i=1}^N
\sum_{d=1}^{|\nu|} p_{y_id}
\pi_d= \prod_{k=1}^{|\mathcal{P}|} \Biggl(
\sum_{d=1}^{|\nu|} p_{kd}
\pi_d \Biggr)^{x_k}, \label{eq:lhood_varknown}
\end{equation}
where $x_k= \sum_{i=1}^{N} \mathrm{I}(y_i=k)$ is the number of reads
following exon path $k$.
Equation~(\ref{eq:lhood_varknown}) is log-concave, which guarantees the
existence of a single maximum.
Log-concavity is given by (i) the log function being concave and
monotone increasing,
(ii) $\sum_{d=1}^{|\nu|} p_{kd} \pi_d$ being linear and therefore concave,
and (iii) the fact that a composition $g \circ f$ where $g$ is concave
and monotone increasing
and $f$ is concave is again concave.
To see (iii), notice that
\begin{eqnarray*}
g \circ f\bigl(t \mathbf{z}_1 + (1-t) \mathbf{z}_2
\bigr)& \geq& g\bigl(tf(\mathbf{z}_1) + (1-t)f(\mathbf{z}_2)
\bigr) \\
&\geq& t g \circ f(\mathbf{z}_1) + (1-t) g \circ f({\mathbf
z}_2),
\end{eqnarray*}
where the first inequality is given by $g$ being increasing and $f$ concave,
and the second inequality is given by $g$ being concave.

We complete the probability model with a Dirichlet prior on $\bpi$:
%
%
\begin{equation}
\bpi| \bnu\sim\operatorname{Dir}(q_1,\ldots,q_{|\nu|}).
\label{eq:priorpi}
\end{equation}
In Section~\ref{sec:results} we assess several choices for $q_d$.
By default we set the fairly uninformative values $q_d=2$,
as these induce negligible bias and stabilize the posterior mode
by pooling it away from the boundaries 0 and 1.
It is easy to see that (\ref{eq:priorpi}) is log-concave when $q_d \geq
1$ for all $d$.
Given that (\ref{eq:lhood_varknown}) is also log-concave,
this choice of $\mathbf{q}$ guarantees the posterior to be log-concave,
and therefore the uniqueness of the posterior mode.

\subsection{Fragment length and read start distribution estimates}
\label{ssec:distrib}

Evaluating the exon path probabilities in (\ref{eq:pathprobs}) that
appear in the likelihood (\ref{eq:lhood_varknown}) requires the
fragment start distribution $P_S$ and fragment length distribution $P_L$.
Given that it is not possible to estimate $(P_L,P_S)$ with precision
for each individual gene,
we assume they are shared across multiple genes
(restricting fragments to be no longer than the variant they originated from).
By default we assume that $(P_L,P_S)$ are common across all genes,
but we also studied posing separate distributions
according to gene length.
Supplementary Section~1 shows experimental evidence that,
while $P_L$ remains essentially constant, $P_S$ can depend on gene
length and the experimental setup.
While this option is implemented in our R package,
to allow a direct comparison with previous approaches
here, we assumed a common $(P_L,P_S)$.

Denoting by $T$ the length of variant
$\delta_i$ (in bp), we let $P_L(l \mid\delta)= P_L(l \mid T)=
P(L=l)\mathrm{I}(l \leq T)/P(l \leq T)$. That is, the conditional
distribution of $L$ given $\delta$ is simply a truncated version of
the marginal distribution.

Further,
we assume a common fragment start distribution relative to the variant
length $T$.
Conditional on $L$ and $T$, $P_S$ is truncated so that the fragment
ends before
the end of the variant.
Specifically,
%
%
\begin{eqnarray}\label{eq:startdistrib}
P_S(S \leq s \mid\delta_i, L=l)&=& P \biggl(
\frac{S}{T} \leq z | T,L=l \biggr)
\nonumber
\\[-8pt]
\\[-8pt]
\nonumber
&=& \frac{\varphi ( \operatorname{min}\{ z, {(T-l+1)}/{T} \}  )}{\varphi
( {(T-l+1)}/{T}  )},
\end{eqnarray}
where $z=s/T$ and $\varphi(z)=P(\frac{S}{T} \leq z)$ is the
distribution of the relative read start $\frac{S}{T}$.

To estimate $P_{L}$ note that the fragment
length is unknown for fragments that span multiple exons,
but it is known
exactly when both ends fall in the same exon.
Therefore, we select all such fragments and estimate $P_L$ with the
empirical probability mass function of the observed fragment lengths.
In order to prevent short exons from inducing a selection bias,
we only use exons that are substantially longer than
the expected maximum fragment length (by default $>1000$~bp).

Estimating the fragment start distribution $P_S$ is more challenging,
as we do not know the variant that generated each fragment and
therefore its relative start position cannot be determined. We address this
issue by selecting genes that have a single annotated variant, as, in
principle, for these genes all fragments should have been generated by
that variant. Of course, the annotated genome does not contain all
variants and, therefore, a proportion of the selected fragments may not
have been generated by the assumed variant.
However, the annotations
are expected to contain most common variants ({i.e.}, with highest
expression)
and, hence, most of the selected fragments should correspond to the
annotated variant. Under this assumption, we can determine the exact start
$S_i$ and length $L_i$ for all selected fragments.
A difficulty in estimating the read start
distribution is that the observed $(S_i,L_i)$ pairs are truncated so
that $S_i+L_i<T$, whereas we require the untruncated cumulative
distribution function $\rho(\cdot)$ in (\ref{eq:startdistrib}).
Fortunately, the truncation point for each $(S_i,L_i)$ is known and,
therefore, one can simply obtain a Kaplan--Meier estimate of
$\rho(\cdot)$ [\citet{kaplan:1958}]. We use the
function \texttt{survfit} in the R \texttt{survival} package
[\citet{rpkg:survival}].

\section{Model fitting}
\label{sec:fitting}

We provide algorithms to obtain a point estimate for $\bpi$,
asymptotic credibility intervals and posterior samples.

Following a 0--1 loss, as a point estimate we report the posterior mode,
which is obtained by maximizing the product of (\ref
{eq:lhood_varknown}) and (\ref{eq:priorpi}),
subject to the constraint $\sum_{d=1}^{|\nu|} \pi_d= 1$.
We note that maximum likelihood estimates are obtained by simply
setting $q_d=1$ in (\ref{eq:priorpi}).
This constrained optimization can be performed with many numerical
optimization algorithms.
Here we used the EM algorithm [\citet{dempster:1977}],
as it is computationally efficient
even when the number of variants $|\nu|$ is large.
For a detailed derivation see Appendix \ref{sec:emalgorithm}.
As noted above, for $q_d>1$ the log-posterior is concave
and, therefore, the algorithm converges to the single maximum.
The steps required for the algorithm are as follows:
\begin{longlist}[1.]
\item[1.] Initialize $\pi_d^{(0)}= q_d / \sum_{d=1}^{|\nu|} q_d$.
\item[2.] At iteration $j$, update
$\pi_d^{(j+1)} = q_d-1+ \sum_{k=1}^{|\mathcal{P}|} x_k \frac{p_{kd} \pi
_d^{(j)}}{\sum_{i=1}^{|\nu|} p_{ki} \pi_i^{(j)}}$.
\end{longlist}
Step 2 is repeated until the estimates stabilize.
In our examples we required $| \pi_d^{(j+1)} - \pi_d^j | < 10^{-5}$ for
all $d$.
Notice that $p_{kd}$ and $x_k$ remain constant through all iterations
and, hence, they need to be computed only once.

We characterize the posterior uncertainty asymptotically
using a normal approximation in the re-parameterized space
$\theta_d= \operatorname{log}  ( \pi_{d+1}/\pi_1  )$, $d=1,\ldots,|\nu|-1$
and the delta method [\citet{casella:2001}].
Denote by $\bolds{\mu}$ the posterior mode for $\bolds{\theta}=(\theta
_1,\ldots,\theta_{|\nu|-1})$
and by $S$ the Hessian of the log-posterior evaluated at $\bolds{\theta
}=\bolds{\mu}$.
Further, let $\pi(\bolds{\theta})$ be the inverse transformation
and $G(\bolds{\theta})$ the matrix with $(d,l)$ element
$G_{dl}= \frac{\partial}{\partial\theta_l} \pi_d(\bolds{\theta})$.
Detailed expressions for $S$, $\pi(\bolds{\theta})$ and $G(\bolds{\theta})$
are provided in Appendix \ref{sec:postapprox_modelpars}.
The posterior for $\bolds{\theta}$ can be asymptotically approximated by
$N(\bolds{\mu},\Sigma)$, where $\Sigma= S^{-1}$.
Hence, the delta method approximates the
posterior for $\bpi$ with $N  (\pi(\bolds{\mu}),G(\bolds{\mu})' S
G(\bolds{\mu})  )$.

The asymptotic approximation is also useful for the
following independent proposal Metropolis--Hastings scheme.
Initialize $\theta^{(0)} \sim T_3(\bolds{\mu},\Sigma)$
and notice that a prior $P_{\pi}(\bpi)$ on $\bpi$ induces a prior
$P_{\theta}(\bolds{\theta})= P_{\pi}(\bpi(\bolds{\theta})) \times|G(\bolds{\theta})|$
on $\bolds{\theta}$, where $G(\bolds{\theta})$ is as above.
At iteration $j$, perform the following steps:
\begin{longlist}[1.]
\item[1.] Propose $\bolds{\theta}^* \sim T_3(\bolds{\mu},\Sigma)$
and let $\bpi^*= \bolds{\pi}(\bolds{\theta}^*)$.
\item[2.] Set $\bolds{\theta}^{(j)}= \bolds{\theta}^*$ with probability $\operatorname
{min}\{1,\lambda\}$, where
%
%
\begin{equation}
\lambda= \frac{P(\mathbf{Y}| \bpi^*,\bnu) P_{\pi}(\bpi^*) |G(\bolds{\theta}^*)|} {
P(\mathbf{Y}| \bpi^{(j-1)},\bnu) P_{\pi}(\bpi^{(j-1)}) |G(\bolds{\theta}^{(j-1)})|} \frac{T_3(\bolds{\theta}^{(j-1)}; \bolds{\mu}, \Sigma) }{T_3(\bolds{\theta}^*;
\bolds{\mu}, \Sigma)}. \label{eq:mhaccept_knownvar}
\end{equation}
Otherwise, set $\bolds{\theta}^{(j)}= \bolds{\theta} ^{(j-1)}$.
\end{longlist}
Posterior samples can be obtained by
discarding some burn-in samples and repeating the
process until practical convergence is achieved.
By default we suggest 10,000 samples with a 1000 burn-in,
as it provided sufficiently high numerical accuracy
when comparing two independent chains
(Supplementary Section~2).

\section{Results}
\label{sec:results}

We assess the performance of our approach in simulations
and two experimental data sets.
We obtained the two human sample K562
replicates\footnote{\href{ftp://ftp.sanger.ac.uk/pub/gencode/rgasp/RGASP1/inputdata/human_fastq/}{ftp://ftp.sanger.ac.uk/pub/gencode/rgasp/RGASP1/inputdata/human\_fastq/}.}
from the RGASP project (\href{http://www.gencodegenes.org/rgasp}{www.gencodegenes.org/rgasp})
and two 
\citet{encode} replicated samples obtained from
A549 cells
(accession number
wgEncodeEH002625\footnote{\href{http://genome.ucsc.edu/ENCODE}{genome.ucsc.edu/ENCODE}.}).
We compare our
results with Cufflinks [\citet{trapnell:2012}],
FluxCapacitor [\citet{montgomery:2010}] and BitSeq [\citet{glaus:2012}].
Cufflinks is based on a probabilistic model akin to Casper,
but uses exon and exon junction counts instead of full exon paths,
assumes that fragment lengths are normally distributed
and estimates the read start distribution in an iterative manner.
FluxCapacitor is also based on exon and exon junction counts,
but uses a method of moments type estimator. 
BitSeq uses a full Bayesian model at the base-pair resolution
({i.e.}, data is not summarized as counts)
and estimates the read start distribution with a two-step procedure.

Regarding sequence alignment, for Casper, Cufflinks and FluxCapacitor
we used TopHat [\citet{trapnell:2009}] with the human genome hg19,
using the default parameters and a 200~bp average insert size.
BitSeq required aligning to the transcriptome with Bowtie [\citet{soft:bowtie}].

\subsection{Simulation study}
\label{ssec:simulation}

\begin{figure}

\includegraphics{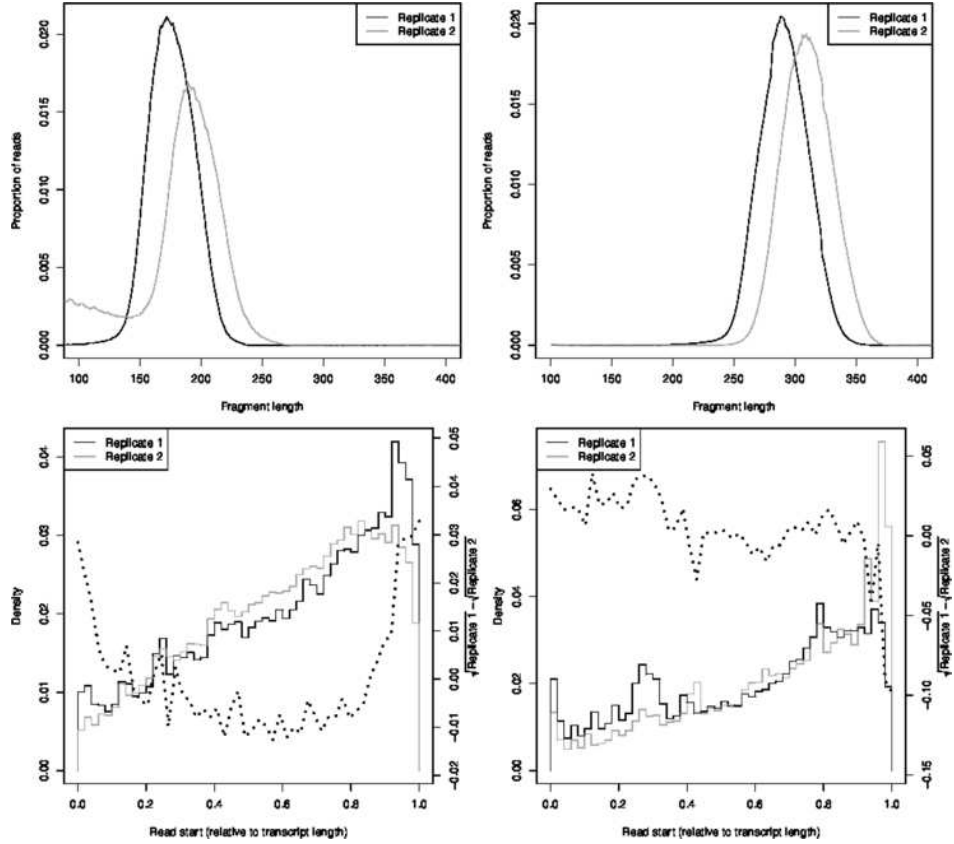}

\caption{Estimated fragment length (top) and start (bottom)
distributions in K562 data (left) and A549 data (right).
Black dotted line: difference in $\sqrt{P_S}$ between replicates
(values in secondary y-axis).}
\label{fig:distribestimates}
\end{figure}

We generated human genome-wide RNA-seq data, setting the simulations
to resemble the K562 RGASP data in order to keep them as realistic as
possible.
Figure~\ref{fig:distribestimates} (left) shows our estimates $\hat
{P}_S$ and $\hat{P}_L$.
We set $P_S$ and $\bpi$ for each gene with 2 or more variants to their
estimates in the K562 data.
For each gene we simulated a number of fragments 
equal to that observed in the K562 sample.

We considered a Casper-based and a Cufflinks-based simulation scenario.
In the former we set $\bpi$ and $P_L$ to the Casper estimates ($q_d=2$).
The second scenario was designed to favor Cufflinks
by using its $\bpi$ estimates and setting $P_L$ to its assumed Normal
distribution
(mean${}={}$200, standard deviation${}={}$20).
We indicated the data-generating $P_L$ to Cufflinks,
whereas the remaining methods estimated it from the data.
An important difference between scenarios is that Casper estimates with $q_d=2$
are pooled away from the boundary, hence, $\pi_d$ is never exactly 0 or 1,
whereas the Cufflinks estimates were often in the boundary
(Supplementary Figure~4).
Genes with less than 10 reads per kilobase per million (RPKM)
were excluded from all calculations to reduce biases due to low expression.

\begin{table}
\caption{Mean absolute and square errors, bias and variance for
simulation study for Casper (top) and Cufflinks estimates (bottom)}
\label{table:varbias}
\begin{tabular*}{\textwidth}{@{\extracolsep{\fill}}lcccc@{}}
\hline
& \textbf{MAE} & \textbf{MSE} & \textbf{Bias sqrt} & \textbf{Variance} \\
\hline
\multicolumn{5}{c}{{Casper-based simulations}}\\
Casper ($q_d=1$) & 0.094 & 0.028 & 0.004 & 0.024 \\
Casper ($q_d=2$) & 0.055 & 0.004 & 0.004 & 0.004 \\
Cufflinks & 0.141 & 0.050 & 0.028 & 0.022 \\
FluxCapacitor & 0.151 & 0.054 & 0.022 & 0.032 \\ [6pt]
\multicolumn{5}{c}{Cufflinks-based simulations} \\
Casper ($q_d=1$) & 0.100 & 0.055 & 0.021 & 0.034 \\
Casper ($q_d=2$) & 0.111 & 0.035 & 0.032 & 0.003 \\
Cufflinks & 0.127 & 0.073 & 0.045 & 0.028 \\
FluxCapacitor & 0.138 & 0.078 & 0.036 & 0.042 \\
\hline
\end{tabular*}
\end{table}

We estimated $\bpi$ from the simulated data using our approach with prior
parameters $q_d=1$ and $q_d=2$, Cufflinks and FluxCapacitor.
Table~\ref{table:varbias} reports the absolute and square errors
($|\pi_d - \hat{\pi}_d|$ and $(\pi_d - \hat{\pi}_d)^2$)
averaged across all $18\mbox{,}909$ isoforms and 100 simulated data sets for
both simulation settings.
We also report the average squared bias and variance.
The Cufflinks and FluxCapacitor MAE are over 2.5 and 2.7 folds greater
than that for Casper with $q_d=2$ (1.5 and 1.6 for $q_d=1$, resp.)
in the Casper-based scenario.
In the Cufflinks-based simulation the reductions were 1.14 and 1.24
fold (1.27 and 1.38 for $q_d=1$).
The improvements in MSE are even more pronounced, with an over 2 fold
improvement for $q_d=2$ even in the Cufflinks-based simulation.
Casper also shows a marked improvement in bias for $q_d=1$ and variance
for $q_d=2$.
See Supplementary Figure~4 for corresponding plots.


\begin{figure}

\includegraphics{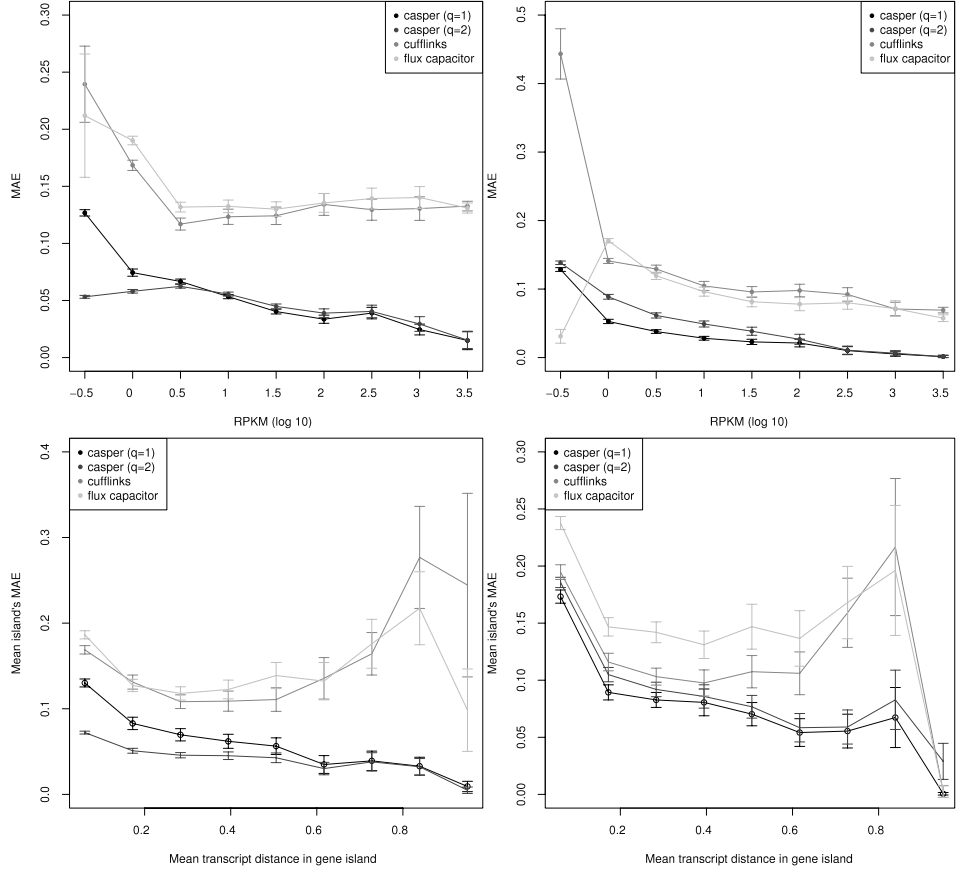}

\caption{Simulation study. Mean absolute error vs. RPKM for Casper
estimates \textup{(a)} and Cufflink estimates \textup{(b)} and the mean base pair
difference between variants in a gene for Casper \textup{(c)} and
Cufflinks-based simulations \textup{(d)}.}
\label{fig:mae}
\end{figure}

Figure~\ref{fig:mae} (top) shows the MAE for each transcript as a
function of
RPKM, a~measure of overall gene expression.
Casper improves the estimates for essentially
all RPKM values in both simulation settings.
Figure~\ref{fig:mae} (bottom) assesses the MAE {vs.} the mean
pairwise difference
between variants in a gene (number of base pairs not shared).
When variants in a gene share most exons this difference is small,
{that is}, variants are harder to distinguish.
Casper estimates are the most accurate at all similarity levels,
with the MAE decreasing as variants become more differentiated.
Interestingly, Cufflinks and FluxCapacitor show lower MAE
as similarity increases from low to medium,
but then MAE becomes higher and more variable in genes
with medium-highly differentiated variants.
These results illustrate the advantage of using full exon paths,
which provide more resolution in assigning reads to splicing variants.

Finally, we assessed the frequentist coverage probabilities for the
asymptotic 95\% credibility intervals
(Section~\ref{sec:fitting}), finding that in 95.04\% of the cases they
contained the true value.


\subsection{Experimental data from RGASP project}
\label{ssec:rgasp}

The two K562 replicates were sequenced in 2009 with Solexa sequencing.
The read length was 75~bp and the mean fragment length indicated in the
documentation
is 200~bp for both replicates.
Figure~\ref{fig:distribestimates} (top, left) shows the estimated
fragment length
distributions.
We observe that the mean length differs significantly from 200~bp
and that there are important differences between replicates.
Replicate 2 shows a heavy left tail that indicates
a subset of fragments substantially shorter than average. This
distributional shape cannot be
captured with the usual parametric distributions.
Figure~\ref{fig:distribestimates} (left, bottom) shows the relative start
distribution.
We observe more sequences located near the transcript end in replicate 1,
{that is}, a higher 3' bias.
The differences between replicates illustrate the need of flexibly
modeling these distributions for each sample separately.
In fact, we found that $\hat{P}_S$ differed across genes with varying length
(Supplementary Section~1 and Supplementary Figure~1),
the 3' bias being stronger in genes shorter than 3 kilo-bases.

\begin{figure}
\centering
\begin{tabular}{@{}c@{}}

\includegraphics{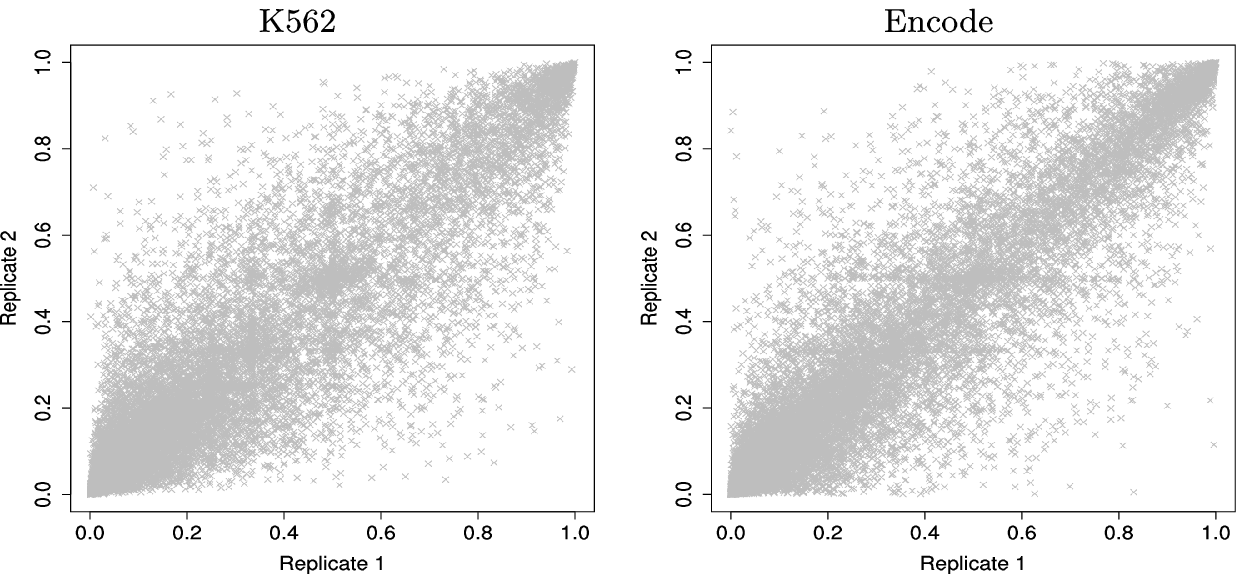}
\\
\footnotesize{(a)}\\[3pt]

\includegraphics{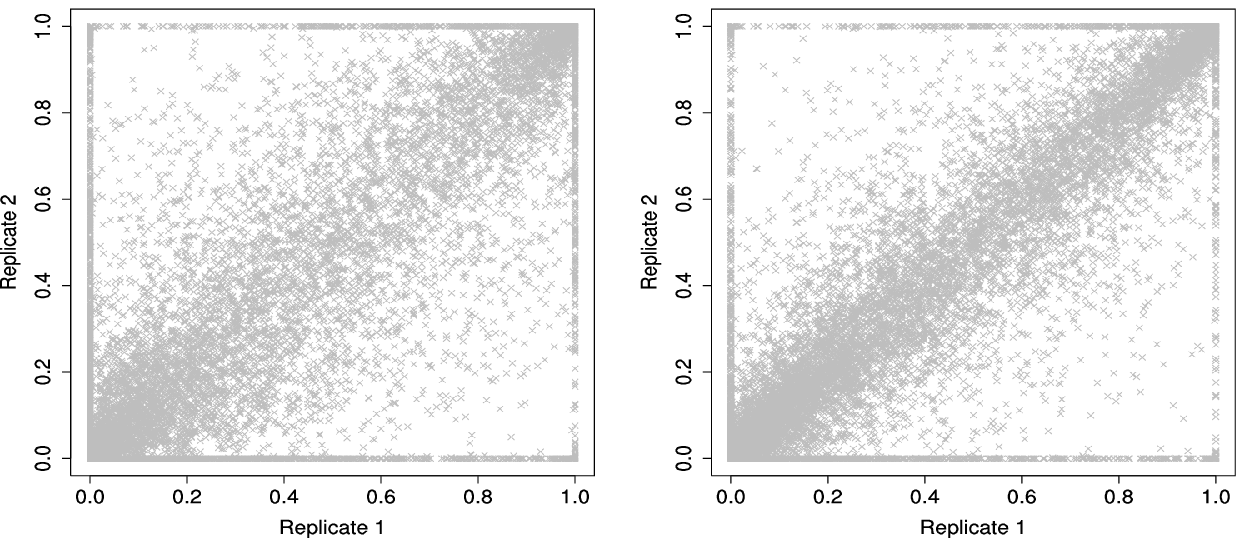}
\\
\footnotesize{(b)}\\[3pt]

\includegraphics{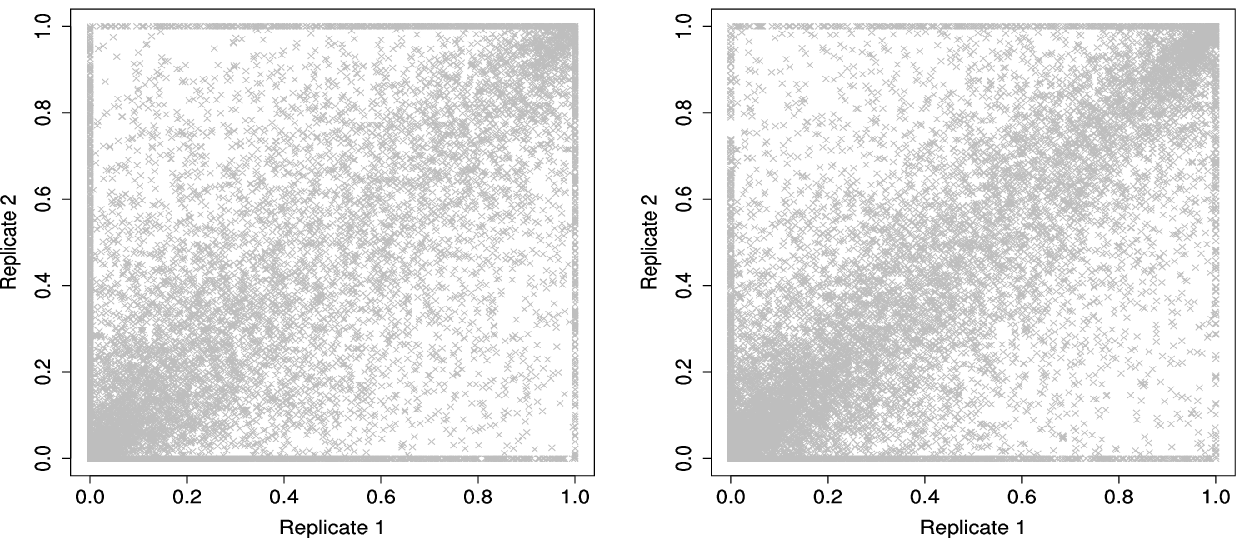}
\\
\footnotesize{(c)}
\end{tabular}
\caption{Comparison of estimated isoform expression $\pi_d$ between two
replicates in K562 and ENCODE studies.
\textup{(a)} Casper with $q_d=2$; \textup{(b)} Cufflinks; \textup{(c)} FluxCapacitor; \textup{(d)} BitSeq.}
\label{fig:k562corr}
\end{figure}

\setcounter{figure}{3}
\begin{figure}
\centering
\begin{tabular}{@{}c@{}}

\includegraphics{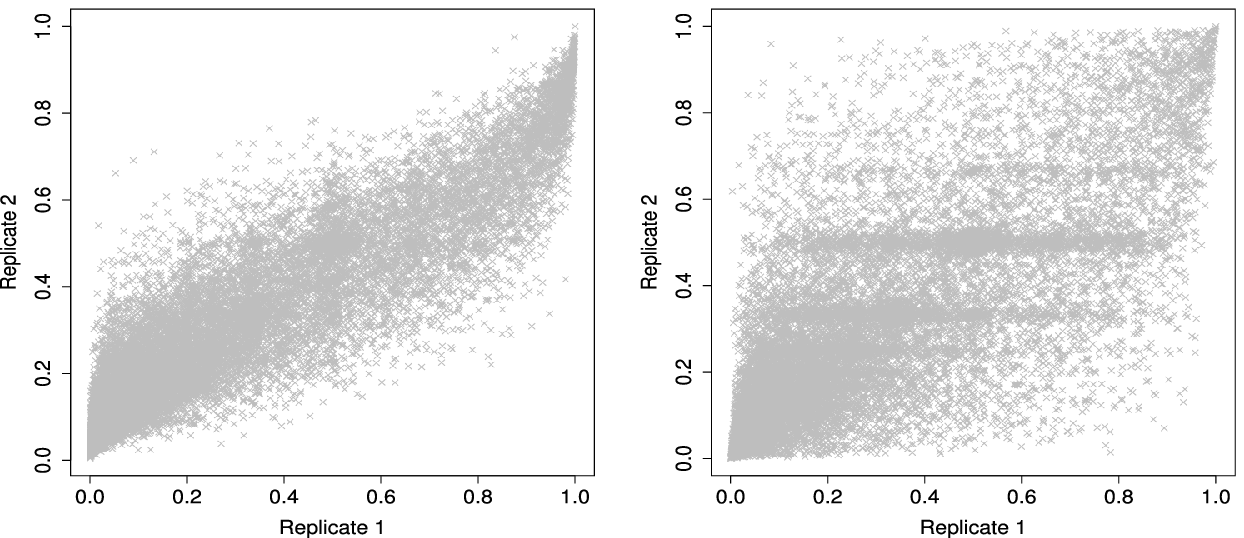}
\\
\footnotesize{(d)}
\end{tabular}
\caption{(Continued).}
\end{figure}

We estimated the expression of human splicing variants
in the UCSC genome version hg19 for the two replicated samples separately.
Figure~\ref{fig:k562corr} (left) and Table~\ref{tab:mad} compare the
estimates obtained in the two
samples.
The Mean Absolute Difference (MAD) between replicates was 0.064 for Casper,
0.126 for Cufflinks (97\% increase), 16.2 for FluxCapacitor (253\%
increase) and 8.5 for BitSeq (31\% increase).
Figure~\ref{fig:k562corr} shows a roughly linear correlation for Casper,
Cufflinks and FluxCapacitor, the latter two frequently providing $\hat
{\pi}_d=0$
in one replicate and $\hat{\pi}_d=1$ in the other.
BitSeq avoids these boundaries but exhibits a strongly nonlinear association.
In terms of computational time,
all methods required roughly 10--20~min on 4 processors.
Because BitSeq models the data at the base-pair resolution,
it required substantially longer time to run on 12 cores.

These results suggest that Casper provides clear advantages
even with earlier sequencing technologies.

\subsection{Experimental data from ENCODE project}
\label{ssec:encode}

The two A549 replicated samples were sequenced in 2012 using Illumina
HiSeq 2000.
The read length was 101~bp and the average fragment length was roughly 300~bp
(Figure~\ref{fig:distribestimates}, top right).
These are substantially longer than the 2009 samples from Section~\ref{ssec:rgasp},
and reflect the improvement in sequencing technologies.
Similar to Section~\ref{ssec:rgasp}, Figure~\ref{fig:distribestimates}
reveals important differences in the fragment length (top, right)
and start (bottom, right) distributions between samples.
See also Supplementary Section~1 and Supplementary Figure~2,
where $\hat{P}_S$ exhibits a stronger 3' bias for genes longer than 5
kilo-bases.

Figure~\ref{fig:k562corr} (left) and Table~\ref{tab:mad} compare the
estimates obtained in the two
replicates.
Similar to the RGASP study (Section~\ref{ssec:rgasp}),
Casper shows a roughly linear association and substantially higher
consistency between replicates.
The MAD between replicates was 0.057 for Casper,
9.0 for Cufflinks (58\% increase), 12.7 for FluxCapacitor (223\% increase)
and 0.098 for BitSeq (72\% increase).
The computational time for Casper was comparable to that of Cufflinks,
higher than FluxCapacitor and substantially lower than BitSeq.
The findings show that the advantage of modeling exon path counts
over pairwise exon connections
remains pronounced as technology evolves to sequence longer fragments.

\begin{table}[b]
\caption{K562 and Encode studies. Mean absolute difference (MAD) in
$\hat{\pi}_d$ between replicates
and CPU time on 2.8~\textup{GHz}, 32~\textup{Gb} OS X computer ($+$: 4 cores; $*$: 12 cores)}
\label{tab:mad}
\begin{tabular*}{\textwidth}{@{\extracolsep{\fill}}lcccc@{}}
\hline
& \multicolumn{2}{c}{\textbf{K562}} & \multicolumn{2}{c@{}}{\textbf{Encode}}
\\[-6pt]
& \multicolumn{2}{c}{\hrulefill} & \multicolumn{2}{c@{}}{\hrulefill} \\
& \textbf{MAD} & \textbf{CPU} & \textbf{MAD} & \textbf{CPU} \\
\hline
Casper$^{+}$ & \phantom{0}$6.4\times10^{-2}$ & 11.1~min & \phantom{0}$5.7\times10^{-2}$ & 2~h
11~min \\
Cufflinks$^{+}$ & $12.6\times10^{-2}$ & 21.4~min & \phantom{0}$9.0\times10^{-2}$
& 2h 13~min \\
Flux$^{+}$ & $16.2\times10^{-2}$ & 9.0~min & $12.7\times10^{-2}$ & 1~h
17~min \\
BitSeq$^{*}$ & \phantom{0}$8.5\times10^{-2}$ & 1~day 13~h & \phantom{0}$9.8\times10^{-2}$ &
8~h 40~min \\
\hline
\end{tabular*}
\end{table}

We now consider the possibility that some expressed transcripts may not be
present in the UCSC genome annotations.
We used a Cufflinks RABT module to identify novel transcripts,
and then run Casper to jointly estimate their expression with UCSC transcripts.
Cufflinks-RABT found
12,512 gene islands with no new transcripts, 6229 with some new
transcripts and 1527 completely new genes in sample 1.
For sample 2 the figures were 11,912, 6983 and 1378 completely new genes.
While new transcripts had negligible influence on genes with no new transcripts,
in the remaining genes $\hat{\pi}_d$ decreased so that a proportion of
the expression could be assigned to the new variants.
For further details see Supplementary Section~4.
These findings suggest that current genome annotations may miss a
nonnegligible number
of expressed variants.
For a careful assessment we recommend following a strategy akin to ours here,
{that is}, combining our approach with a \textit{de novo} transcript
discovery method.

\section{Discussion}
\label{sec:discussion}

We proposed a model to estimate the expression of a set of known
alternatively spliced variants from RNA-seq data.
The model improves upon previous
proposals by using exon paths, which are more informative than single
or pairwise exon counts, and by flexibly estimating the fragment start
and length distributions. We provided
computationally efficient algorithms for obtaining point estimates,
asymptotic credibility intervals and posterior samples.

We found that a fairly uninformative prior with $q_d=2$ improves precision
relative to the typical $q_d=1$
equivalent to maximum likelihood estimation. The advantages stem
from the usual shrinkage argument: $q_d=2$ pools
the estimates away from the boundaries and reduces variance.
Compared to competing approaches, we
observed substantial MSE reductions in simulations
and increased correlation between experimental replicates.
In modern studies we found that roughly 2 sequences out of 3 visited
$>2$ exon regions distinguishing variants,
suggesting that the current standard of reporting
pairwise exon junctions adopted by most public databases is far from optimal.
Reporting exon paths would allow researchers to estimate isoform expression
at a much higher precision.
Given that the methodology is implemented in the R package \texttt{casper}, we believe
that it should be of great value to practitioners.

\begin{appendix}\label{app}
\section{Derivation of exon path probabilities}
\label{sec:path_prob}

Here we describe how to compute the probability $p_{kd}$ of observing
exon path $k$
for any splicing variant $d$.
Equivalently, we denote $d$ by $\bolds{\delta}=(i_1,\ldots,i_{|\delta|})$,
where $i_j$ indicates the $j${th} exon within $d$.
Consider variant $\bolds{\delta}$
after splicing, {that is}, after removing the introns.
The new exon start positions are given by $s_1^*=1$
and $s_{k+1}^*= s_k^* + e_{i_k} - s_{i_k} +1$
for $k=1,\ldots,|\delta|-1$.
The end of exon $k$ is $s_{k+1}^*-1$.
Denote by $S$ the read start position, $L$ the fragment length,
$r$ the read length,
and let $T=s_{|\delta|}^*-1$ be the transcript length of $\bolds{\delta}$.

The goal is to compute
$P(\bolds{\iota}_l=(i_j,\ldots,i_{j+k}), \bolds{\iota}_r=(i_{j'},\ldots
,i_{j'+k'})| \bolds{\delta})$.
We note that both $i_j,\ldots,i_{j+k}$ and $i_{j'},\ldots,i_{j'+k'}$
must be consecutive exons under variant~$\bolds{\delta}$,
otherwise the probability of the path is zero.
The left read follows the exon path $\bolds{\iota}_l=(i_j,\ldots,i_{j+k})$
if and only if the read:
\begin{longlist}[1.]
\item[1.] Starts in exon $j$, {that is}, $s_j^* \leq S \leq s_{j+1}^*-1$.
\item[2.] Ends in exon $j+k$, {that is}, $s_{j+k}^* \leq S +r -1 \leq
s_{j+k+1}^*-1$.
\end{longlist}
Similarly, the right read follows $\bolds{\iota}_r=(i_{j'},\ldots,i_{j'+k'})$
if and only if $s_{j'}^* \leq S + L -r \leq s_{j'+1}-1$
and $s_{j'+k'}^* \leq S + L -1 \leq s_{j'+k'+1}^*-1$.
This implies that
the desired probability
can be written as $P(a_1 \leq S \leq b_1, a_2 \leq S + L \leq b_2 | \bolds{\delta})$, where
%
%
\begin{eqnarray}\label{eq:path_constants}
a_1&= &\operatorname{max}\bigl\{s_j^*,s_{j+k}^*-r+1
\bigr\},
\nonumber
\\
b_1&= &\operatorname{min}\bigl\{s_{j+1}^*-1,
s_{j+k+1}^* -r \bigr\},
\nonumber
\\[-8pt]
\\[-8pt]
\nonumber
a_2&= &\operatorname{max}\bigl\{s_{j'}^*+r,
s_{j'+k'}^* +1 \bigr\},
\\
b_2&= &\operatorname{min}\bigl\{s_{j'+1}^*+r-1,
s_{j'+k'+1}^* \bigr\}.\nonumber
\end{eqnarray}

Assuming that the distribution of $(S,L)$ depends on $\bolds{\delta}$ only
through its transcript length $T$,
we can write $P(a_1 \leq S \leq b_1, a_2 \leq S + L \leq b_2 | T)=$
%
%
\begin{eqnarray}\label{eq:prob_path}
&&\sum_{l} P(a_1 \leq S \leq
b_1, a_2 \leq S + L \leq b_2 | T, L=l)
P(L=l|T)
\nonumber
\\[-8pt]
\\[-8pt]
\nonumber
&&\qquad=\sum_{l} P\bigl(\operatorname{max}
\{a_1,a_2-L\} \leq S \leq\operatorname{min}\{
b_1, b_2-L \} | T, L=l\bigr) P(L=l|T).
\end{eqnarray}
In order to evaluate (\ref{eq:prob_path}), we need to estimate the
fragment length distribution $P(L=l|T)$
and the distribution of the read start position $S$ given $L$.
We assume that $P(L|T)= P(L=l)\mathrm{I}(l \leq T)/P(L \leq T)$,
{that is}, the conditional distribution of $L$ given $T$ is simply
a truncated version of the marginal distribution.
Further, notice that the fragment end must happen before the end of the
transcript, {that is}, $S+L-1 \leq T$
or, equivalently, the relative start position is truncated $S/T \leq
S_T= (T-L+1)/T$.
The relative start distribution is therefore truncated,
{that is},
$P(\frac{S}{T} \leq z | T,L=l)= \frac{\varphi ( \operatorname{min}\{ z, S_T
\}  )}{\varphi ( S_T  )}$,
where $\varphi(z)=P(\frac{S}{T} \leq z)$ is the distribution of the
relative read start $\frac{S}{T}$.

Under these assumptions,
the probability of observing the exon path
$\bolds{\iota}_l=(i_j,\ldots,i_{j+k})$,
$\bolds{\iota}_r=(i_{j'},\ldots,i_{j'+k'})$
under variant $\bolds{\delta}$ is equal to
\begin{eqnarray*}\label{eq:prob_path_final}
&&\sum_l \bigl[ \bigl(\varphi \bigl( \operatorname{min}\bigl\{{b_1}/{T},
{(b_2-l)}/{T}, S_T\bigr\}  \bigr)
\\
&&\hspace*{21pt}{}- \varphi \bigl( \operatorname{min}  \bigl\{ \operatorname{max}\bigl\{ {(a_1-1)}/{T},
{(a_2-l-1)}/{T} \bigr\}, S_T  \bigr\}  \bigr)\bigr)/ {
\varphi ( S_T  )}
 \bigr]_{+} P(L=l|T),
\end{eqnarray*}
where $a_1$, $b_1$, $a_2$ and $b_2$ are given in (\ref{eq:path_constants}).
Given that highly precise estimates of $P(L=l)$ and $\varphi(\cdot)$
are typically available,
for computational simplicity we treat them as known
and plug them into (\ref{eq:prob_path_final}).

\section{EM algorithm derivation}
\label{sec:emalgorithm}

\begin{longlist}[1.]
\item[1.]\textit{E-step}.

Let $\delta_i \in\{ 1,\ldots,|\nu| \}$ be latent variables indicating
the variant that reads $i=1,\ldots,N$ come from.
The augmented log-posterior is proportional to
%
%
\begin{eqnarray}\label{eq:augmented_logl}
l_0(\bpi|\mathbf{y},\bolds{\delta})&=& \operatorname{log} P(\bpi|
\bnu) + \operatorname{log} P(\mathbf{y},\bolds{\delta}| \bpi)
\nonumber
\\[-8pt]
\\[-8pt]
\nonumber
& =& \sum
_{d=1}^{|\nu|} (q_d-1) \operatorname{log}(
\pi_d)
+\sum_{i=1}^N \sum
_{d=1}^{|\nu|} \mathrm{I}(\delta_i=d) \bigl[
\operatorname {log}(p_{y_id}) + \operatorname{log}(\pi_d)
\bigr].
\end{eqnarray}
Considering $\delta_i$ as a random variable, the expected value of (\ref
{eq:augmented_logl})
given $\mathbf{y}$ and $\bpi=\bpi^{(j)}$ is equal to
%
%
\begin{eqnarray}\label{eq:eaugmented_logl}
&&E \bigl( l_0\bigl(\bpi'|\mathbf{y},\bolds{\delta}
\bigr) | \mathbf{y}, \bpi^{(j)} \bigr)
\nonumber
\\
&&\qquad= \sum
_{d=1}^{|\nu|} (q_d-1) \operatorname{log}(
\pi_d)
\\
&&\qquad\quad{}+\sum_{i=1}^{N} \sum
_{d=1}^{|\nu|} P\bigl(\delta_i=d|
y_i, \bpi^{(j)}\bigr) \bigl( \operatorname{log}(p_{y_id})
+ \operatorname{log}\bigl(\pi_d'\bigr) \bigr).\nonumber
\end{eqnarray}

\item[2.]\textit{M-step}.

The goal is to maximize (\ref{eq:eaugmented_logl}) with respect to $\bpi'$.
Let $\gamma_{id}= P(\delta_i=d| y_i, \bpi^{(j)})$ and
re-parameterize $\pi_{|\nu|}= 1 - \sum_{d=1}^{|\nu|-1} \pi_d$.
Setting the partial derivatives with respect to $\pi_d'$ to zero gives
the system of equations
\[
\frac{\pi_d'}{1 - \sum_{d=1}^{|\nu|-1} \pi_d'}= \frac{q_d-1+ \sum_{i=1}^N \gamma_{id}}{q_{|\nu|}-1+ \sum_{i=1}^N \gamma_{i|\nu|}},
\]
which has the trivial solution $\pi_d' \propto q_d-1+ \sum_{i=1}^N
\gamma_{id}$.
By plugging in
$\gamma_{id}= p_{y_id} \pi_d^{(j)} / \sum_{d=1}^{|\nu|} p_{y_id} \pi_d^{(j)}$,
we obtain
\[
\pi_d' \propto q_d-1+ \sum
_{i=1}^N \frac{p_{y_id} \pi_d^{(j)}}{\sum_{d=1}^{|\nu|} p_{y_id} \pi_d^{(j)}}.
\]
Finally, since $x_k= \sum_{i=1}^N \mathrm{I}(y_i=k)$, we can group all
$y_i$'s taking the same value and find the maximum as
%
%
\begin{equation}
\pi_d' \propto q_d-1+ \sum
_{k=1}^{|\mathcal{P}|} x_k \frac{p_{kd} \pi
_d^{(j)}}{\sum_{d=1}^{|\nu|} p_{kd} \pi_d^{(j)}},
\label{eq:mstepsol}
\end{equation}
re-normalizing $\bpi'$ so that $\sum_{d=1}^{|\nu|} \pi_d'= 1$.
\end{longlist}

\section{Asymptotic posterior approximation}
\label{sec:postapprox_modelpars}

Here we derive an asymptotic approximation to
$P(\bpi\mid\bnu, \mathbf{Y})$,
the posterior distribution of
the splicing variants expression $\bpi$ conditional on a model $\bnu$
and the observed data $\mathbf{Y}$.
Given that $\bpi=(\pi_1,\ldots,\pi_{|\nu|}) \in [ 0,1  ]^{|\nu|}$
with $\sum_{i=1}^{|\nu|} \pi_i= 1$,
we re-parameterize to
$\bolds{\theta}=(\theta_1,\ldots,\theta_{|\nu|-1}) \in\Re^{|\nu|-1}$,
where
$\theta_d= \operatorname{log}  ( \frac{\pi_{d+1}}{\pi_1}  )$ for
$d=1,\ldots,|\nu|-1$.
The goal is to approximate
$P(\bolds{\theta} \mid\bnu, \mathbf{Y}) \sim N(\bolds{\mu}, \Sigma)$.
For notational simplicity, in the remainder of the section
we drop the conditioning on $\bnu$.


A prior $P_{\pi}(\bpi)$ induces a prior
$P_{\theta}(\bolds{\theta})= P_{\pi}(\bpi(\bolds{\theta})) \times|G(\bolds{\theta})|$
on $\bolds{\theta}$,
where
$G(\bolds{\theta})$ is the matrix with $(d,l)$ element $G_{dl}= \frac
{\partial}{\partial\theta_l} \pi_d(\bolds{\theta})$
and
inverse transform $\pi_1(\bolds{\theta})=  ( 1 + \sum_{j=1}^{|\nu|-1}
e^{\theta_j}  )^{-1}$,
$\pi_d(\bolds{\theta})= \pi_1(\bolds{\theta}) \operatorname{exp}\{ \theta_{d-1} \}$
for $d>1$.

Define
$f(\bolds{\theta})= \operatorname{log}  ( P(\mathbf{Y} | \bolds{\theta})  ) +
\operatorname{log}  ( P_{\theta}(\bolds{\theta})  )$.
Up to an additive constant, $f(\bolds{\theta})$ is equal to the target
log-posterior distribution of $\bolds{\theta}$ given $\mathbf{Y}$.
We center the approximating Normal at the posterior mode,
{that is}, $\bolds{\mu} = \operatorname{argmax}_{\bolds{\theta} \in\Re^{|\nu
|-1}} f(\bolds{\theta})$.
We set $\Sigma=S^{-1}$, where $S$ is the Hessian of $f(\bolds{\theta})$
evaluated at $\bolds{\theta}= \bolds{\mu}$
with $(l,m)$ element $S_{lm}= \frac{\partial^2}{\partial\theta_l\,
\partial\theta_m} f(\bolds{\theta})$.
We approximate
$\mu_d= \operatorname{log}  ( \frac{\pi_{d+1}^*}{\pi_d}  )$,
where $\bpi^*$ is the posterior mode for $\bpi$
provided by the EM algorithm.

Under a $\bpi\sim\operatorname{Dirichlet}(\mathbf{q})$ prior,
simple algebra gives
$\sigma_{lm}= \frac{\partial^2}{\partial\theta_l\, \partial\theta_m}
f(\bolds{\theta})=$
%
%
\begin{eqnarray}\label{eq:hessian_joint}
\quad&&\sum_{k=1}^{|\mathcal{P}|} x_k
\frac{  ( \sum_{d=1}^{|\nu|} p_{kd}
H_{dlm}  )  ( \sum_{d=1}^{|\nu|} p_{kd} \pi_d(\bolds{\theta})
) -  ( \sum_{d=1}^{|\nu|} p_{kd} G_{dl}  )  ( \sum_{d=1}^{|\nu|} p_{kd} G_{dm}  )} {
( \sum_{d=1}^{|\nu|} p_{kd} \pi_d(\bolds{\theta})  )^2}
\nonumber
\\[-8pt]
\\[-8pt]
\nonumber
\quad&&\qquad{}
+ \sum_{d=1}^{|\nu|} (q_d -1)
\frac{H_{dlm} \pi_d(\bolds{\theta}) - G_{dl} G_{dm} }{\pi_d(\bolds{\theta})^2},
\end{eqnarray}
where
$x_k=\sum_{i=1}^{N} \mathrm{I}(y_i=k)$ is the number of reads following
exon path $k$,
$p_{kd}=P(Y_i=k \mid\delta=d)$ is the probability of observing path
$k$ under variant $d$,
the gradient for $\pi_d(\bolds{\theta})$
is $G_{dl}= \frac{\partial}{\partial\theta_l} \pi_d(\bolds{\theta})$
as before and the Hessian is
$H_{dlm}= \frac{\partial^2}{\partial\theta_l\, \partial\theta_m} \pi
_d(\bolds{\theta})$.\looseness=1

We complete the derivation by providing expressions for $G_{dl}$ and $H_{dlm}$.
Let $s(\bolds{\theta})= 1 + \sum_{j=1}^{|\nu|-1} e^{\theta_j}$,
then $G_{dl}=$
%
%
\begin{eqnarray}\label{eq:gradinverse}
&\displaystyle\frac{-e^{\theta_l}}{ s(\bolds{\theta})^2} \qquad\mbox{if } d=1,&
\nonumber
\\[-8pt]
\\[-8pt]
\nonumber
&\displaystyle\frac{-e^{\theta_{d-1} + \theta_l}}{ s(\bolds{\theta})^2} + \mathrm{I}(l=d-1) \frac{e^{\theta_l}}{ s(\bolds{\theta})}\qquad \mbox{if }
d\geq2&
\end{eqnarray}
and $H_{dlm}=$
%
%
\begin{eqnarray}\label{eq:hessinverse}
&\displaystyle\frac{2 e^{\theta_l+\theta_m}}{s(\bolds{\theta})^3} - \mathrm{I}(l=m) \frac
{e^{\theta_l}}{s(\bolds{\theta})^2}\qquad\mbox{if } d=1,&
\\
&\displaystyle\frac{2 e^{\theta_{d-1} + \theta_l + \theta_m}}{s(\bolds{\theta})^3} - \mathrm{I}(l=d-1) \frac{e^{\theta_l +
\theta_m}}{s(\bolds{\theta})}\qquad\mbox{if } d\geq2, m \neq l, m \neq d-1,&
\nonumber\\
&\displaystyle\frac{-2 e^{2\theta_m}}{s(\bolds{\theta})^2} + \frac{2 e^{3 \theta
_m}}{s(\bolds{\theta})^3} + \frac{e^{\theta_m}}{s(\bolds{\theta})} -
\frac
{2e^{2 \theta_m}}{s(\bolds{\theta})^2}\qquad \mbox{if } d\geq2, m=l,
m=d-1,&
\nonumber\\
&\displaystyle\frac{ -e^{\theta_{d-1}+\theta_l}}{ s(\bolds{\theta})^2} + \frac{2
e^{\theta_{d-1}+\theta_l+\theta_m}}{s(\bolds{\theta})^3} \qquad\mbox{otherwise.}&
\nonumber
\end{eqnarray}
\end{appendix}

\section*{Acknowledgments}
D. Rossell and C. Stephan-Otto Attolini contributed equally to this work.
The authors wish to thank Modesto Orozco for useful discussions.

%
\begin{supplement}[id=suppA]
\stitle{Supplementary results}
\slink[doi]{10.1214/13-AOAS687SUPP} 
\sdatatype{.pdf}
\sfilename{aoas687\_supp.pdf}
\sdescription{In \citet{caspersuppl} we assess the dependence of
fragment start and length distributions on gene length, show additional
simulation results, assess MCMC convergence and apply the approach to
transcripts found \textit{de novo}.}
\end{supplement}

%

%


\printaddresses

\end{document}